\documentstyle[12pt]{article}

\def\H{{\cal H}}
\def\h{{\bf h}}
\def\k{{\kappa}}
\def\n{{\nu}}
\def\LL{{\cal L}}
\def\M{{\cal M}}

\def\RR{{\bf R}}
\def\w{{\rm w}}
\def\sa{{\rm sa}}

\def\Im{{\rm Im}\>}
\let\@=\|
\def\?#1?{\left\@#1\right\@}
\def\!#1!{\left\@#1\right\@}
\def\<#1|#2>{\left<#1\vphantom{#2}\right|\left.\vphantom{#1}#2\right>}
\def\|#1|{\left|#1\right|}
\def\(#1){\left(#1\right)}
\def\[#1|#2]{\left\lbrace #1\vphantom{#2}\right|
             \left.\vphantom{#1}#2\right\rbrace}

\begin{document}
\date{October 1994}
\title{Endomorphism Semigroups and Lightlike Translations}
\author{D.~R.~Davidson*\\
Dipartimento di Matematica\\
Universit\`a di Roma ``La Sapienza''$\quad$00185 Roma, Italy\\
Email:  davidson@mat.uniroma1.it
}
\maketitle
\begin{abstract}
Certain criteria are demonstrated for a spatial derivation of a von Neumann
algebra to generate a one-parameter semigroup of endomorphisms of that
algebra.  These are then used to establish a converse to recent results of
Borchers and of Wiesbrock on certain one-parameter semigroups of endomorphisms
of von Neumann algebras (specifically, Type III${}_1$ factors) that appear
as lightlike translations in the theory of algebras of local observables.
\end{abstract}
* This research was partly supported by a fellowship from the
Consiglio Nazionale delle Ricerche.
\newpage

\section{Introduction}

The standard situation for a pair of complementary spacetime regions in the
theory of algebras of local observables, under the assumption of duality in the
vacuum sector, is just that termed standard in the theory of von Neumann
algebras:  we have a von Neumann algebra $\M$ and its commutant $\M'$ acting on
a Hilbert space $\H$, with a common cyclic and separating unit vector $\Omega$,
the vacuum vector.  In the particular situation in which $\M$ and $\M'$
correspond to the observables for a pair of complementary wedge regions (for
definiteness let us take them to be $W_R=\[x|x_1>\|t|]$ and
$W_L=\[x|x_1<-\|t|]$
respectively) it is expected \cite{BW} that the modular automorphism group
$\sigma_t(A)=\Delta^{it}A\Delta^{-it}$ will correspond to the Lorentz velocity
transformations $V_1(2\pi t)$ in the direction orthogonal to the common face
$x_1=t=0$ of the two wedges, and that the modular conjugation $J$ will be a
slight variant of the TCP operator $\Theta$ (so as to give a reflection about
that face).  In that case the lightlike translations
$U(a)=T(a(\hat x_0+\hat x_1))$ will be a strongly continuous one-parameter
group of unitary operators on $\H$, which should have the following
four properties:\\
(a) By Lorentz covariance,
$\Delta^{it}U(a)\Delta^{-it}=U(e^{-2\pi t}a)$ and $JU(a)J=U(-a)$;\\
(b) By the spectral condition,
$U(a)$ should have a positive generator $H$;\\
(c) By isotony,
for $a\geq 0$ the corresponding adjoint action $A\rightarrow U(a)AU(-a)$ should
give a one-parameter semigroup of endomorphisms of $\M$ (and thus for $a\leq 0$
likewise of $\M'$); and, finally,\\
(d) The vacuum vector $\Omega$ should be fixed by all $U(a)$, and thus
annihilated by $H$.

In this connection Borchers has shown \cite{Bo} that these four conditions are
not all independent:  in particular, if the last three hold, then the Lorentz
covariance conditions follow automatically.  Wiesbrock then proved \cite{W}
conversely that if (a), (c), and (d) hold, $U(a)$ automatically has a positive
generator.  In this note we demonstrate that the results of Borchers and of
Wiesbrock are part of a larger chain of converses, and in the process perhaps
shed some further light on these remarkable theorems.  Specifically, we show
that if the generator $H$ gives a derivation $\delta$ of $\M$ satisfying
certain additional conditions, then (a), (b), and (d) imply (c), and in fact
any three of the conditions listed above for $U(a)$ together imply the fourth.
Note that in the local algebra context, it can be shown \cite{Dr} that $\M$ and
$\M'$ must be Type III${}_1$ factors, but this will not be used in the
following; the results will simply be stated in terms of arbitrary von Neumann
algebras.

The situation here is analogous to, but in some respects altogether different
from, the case of spatial derivations that generate automorphism groups of
von Neumann algebras, which has been extensively studied
(\cite{OAQSM}, Section 3.2.5, and references therein).  We will develop the
analogy more specifically after stating Theorem 1, but the obvious relevant
condition is that $U(a)$ should commute with $J$ and with all $\Delta^{it}$;
then the key question is to determine precisely what additional conditions on
the derivation $\delta$ suffice to show that it generates an automorphism
group.  The best result in this direction is that of \cite{BH}, in which the
only additional assumption is that the derivation has a domain $D(\delta)$ such
that $D(\delta)\Omega$ is a core for $H$.  However, the proof of this result is
rather difficult, and does not generalize to the endomorphism case.  We will
make do with more restrictive conditions here, but it would be very interesting
to determine precisely what conditions suffice to guarantee that $\delta$
generates an endomorphism semigroup.  Note that the endomorphism semigroups
studied here are non-pathological counterexamples to the conjecture of
\cite{BR}
(for which many counterexamples are known \cite{BH}).

\section{Endomorphism Semigroups}

If we have a von Neumann algebra $\M$ and its commutant $\M'$ acting on a
Hilbert space $\H$, with a common cyclic and separating vector $\Omega$,
we may define real linear spaces $R=\overline{\M^\sa\Omega}$ and
$R'=\overline{\M^{'\sa}\Omega}$.  Then $\<\psi|\phi>$ is real for all
$\psi\in R,\phi\in R'$, and furthermore $R'$ is precisely the set of all
$\psi$ such that $\<\psi|\phi>$ is real for all $\phi\in R$.  Also,
$D(\Delta^{1/2})=R+iR$ and $D(\Delta^{-1/2})=R'+iR'$ are dense in $\H$,
$R=\[\psi|\psi\in D(\Delta^{1/2}), J\Delta^{1/2}\psi=\psi]$, and
$R'=\[\psi|\psi\in D(\Delta^{-1/2}), J\Delta^{-1/2}\psi=\psi]$.

For any $\psi\in R$, there is a sequence $X_n\in\M^\sa$ such that
$X_n\Omega\rightarrow\psi$, but there need not be a bounded operator
$X\in\M^\sa$ such that $X\Omega=\psi$; in general there is only a closed
symmetric operator $\tilde X$ affiliated with $\M$ such that
$\tilde X\Omega=\psi$, to which the $X_n$ converge on the common core
$\M'\Omega$, so that $\tilde XY\Omega=Y\psi$ for every $Y\in\M'$.  If
$\tilde X$ is self-adjoint, then the $X_n$ will converge to $\tilde X$ in
the strong resolvent sense (\cite{RS}, Theorem VIII.25).

If we are to have $U(a)\M U(-a)\subset\M$ for all $a\geq 0$, then the
generator $H$ of the unitary group $U(a)$ must give a derivation $\delta$
of $\M$ by $\delta(X)=i[H,X]$; however, this derivation will be unbounded,
hence defined only on a dense set, and the problem is to give sufficient
conditions for $\delta$ to generate a semigroup of endomorphisms of $\M$.
Let
\begin{equation}
\M_\epsilon=\[X|U(a)XU(-a)\in\M\,\,\hbox{\rm for all}\,\,
0\leq a\leq\epsilon],
\end{equation}
and let $R_\epsilon=\overline{\M_\epsilon^\sa\Omega}$; then
$\M_\epsilon\supset\M_{\epsilon'}$ and $R_\epsilon\supset R_{\epsilon'}$
whenever $\epsilon'\geq\epsilon$.  In addition, let
\begin{equation}
\M_+=\bigcup_{\epsilon>0}\M_\epsilon\quad\hbox{\rm and}\quad
R_+=\bigcup_{\epsilon>0}R_\epsilon.
\end{equation}
Then $\M_\epsilon$ contains those elements $X$ of $\M$ for which the
differential equation $X(t)'=\delta(X(t))$, $X(0)=X$ in the Banach space $\M$
has a solution curve of length at least $\epsilon$; likewise, $\M_+$ contains
those for which there is a solution curve of any positive length.  Conditions
on $\M_\epsilon$ and $\M_+$ can thus be regarded as local existence conditions
for this differential equation, and it is criteria of this sort that we will
use to control the behavior of the derivation $\delta$.

\bigskip
\noindent{\bf Theorem 1:}~~
{\em
Suppose that $U(a)\Omega=\Omega$ and $U(a)R\subset R$ for all $a\geq 0$, and
that for some $\epsilon>0$, $\Omega$ is cyclic for $\M_\epsilon$, i.e.,
$R_\epsilon+iR_\epsilon$ is dense in $\H$.  Then
$U(a)\M U(-a)\subset\M$ for all $a\geq 0$.
}

\smallskip
\noindent{\bf Proof of Theorem 1:}~~
{
It will suffice to show that $U(a)\M'U(-a)\supset\M'$ for all $a\geq 0$;
we have from our assumptions that $U(a)R'\supset R'$ for all $a\geq 0$.  Let us
pick $a$ such that $0\leq a\leq\epsilon$, so that
$U(a)\M'U(-a)\subset\M_\epsilon'$.  Let $X$ be a self-adjoint element of $\M'$;
then $X\Omega\in R'\subset U(a)R'$, so that there is a sequence $Y_n$ of
self-adjoint elements of $\M'$ such that $U(a)Y_n\Omega\rightarrow X\Omega$.
Now, $X$ and every $X_n=U(a)Y_n U(-a)$ are all in $\M_\epsilon'$, and
$X_n\Omega\rightarrow X\Omega$, so as above the $X_n$ tend to $X$ on the common
core $\M_\epsilon\Omega$.  But the $X_n$ and $X$ are all self-adjoint, so
the $X_n$ tend to $X$ in the strong resolvent sense.  Since
each $X_n\in U(a)\M'U(-a)$, $X$ is affiliated with $U(a)\M'U(-a)$, hence
$X\in U(a)\M'U(-a)$ and $U(a)\M'U(-a)\supset\M'$.  This is so for all
$0\leq a\leq\epsilon$, hence by the semigroup property for all $a\geq 0$.
}
\smallskip

\noindent{\bf Remarks:}~~
{
The analogy between the automorphism and endomorphism cases is now evident:
in the automorphism case, the relevant condition is that $U(a)R=R$ for
all $a\in\RR$; this is equivalent to the commutation of $U(a)$ with $J$
and with all $\Delta^{it}$.  The desired conclusion would then be that
$U(a)\M U(-a)=\M$ for all $a\in\RR$.  Although the situation here is in some
respects similar, there are a number of significant differences.  For example,
if $H$ were positive in the automorphism case, then by the Borchers-Arveson
theorem it would be affiliated with $\M$, but since it annihilates the
separating vector $\Omega$, it would have to vanish.  By contrast, in the
endomorphism case it is possible for $H$ to be positive without being
affiliated with $\M$.  This will occur in the special case of Theorem 2,
in which we are primarily interested, and about which we can say somewhat
more.
}
\bigskip

\noindent{\bf Theorem 2:}~~
{\em
Suppose that $U(a)\Omega=\Omega$ and $U(a)R\subset R$ for all $a\geq 0$, that
$\Delta^{it}U(a)\Delta^{-it}=U(e^{-2\pi t}a)$, and that
$\Omega$ is cyclic for $\M_+$, i.e., $R_++iR_+$ is dense in $\H$.
Then $U(a)\M U(-a)\subset\M$ for all $a\geq 0$.
}
\smallskip

\noindent{\bf Proof of Theorem 2:}~~
{
Notice that $\Delta^{it}R_\epsilon=R_{e^{-2\pi t}\epsilon}$, so that
for any $\epsilon>0$, $R_+=\cup_{t\geq 0}\lbrace\Delta^{it}R_\epsilon\rbrace$.
By assumption, for any $\psi\in\H$, there is some $\epsilon>0$ and some
$\phi\in R_\epsilon+iR_\epsilon$ such that $\<\psi|\phi>\neq 0$.  Thus given
any particular $\epsilon>0$, there is some $\phi\in R_\epsilon+iR_\epsilon$ and
some $t\geq 0$ such that $\<\psi|\Delta^{it}\phi>\neq 0$.  But
$\phi\in R+iR=D(\Delta^{1/2})$, so that $\phi$ is an analytic vector for
$\Delta^{it}$ in the strip $-1/2\leq\Im t\leq 0$.  Thus
$\<\psi|\Delta^{it}\phi>$ is the boundary value of a function analytic in $t$
on that strip, and cannot vanish for all $t\leq 0$.  It follows that
$R_\epsilon+iR_\epsilon=\cup_{t\leq 0}
\lbrace\Delta^{it}(R_\epsilon+iR_\epsilon)\rbrace$ is dense in
$\H$ already, and Theorem 1 applies.
}
\bigskip

The condition that $R_++iR_+$ be dense will be referred to as the local
existence condition of Theorem 2; the condition of Theorem 1 is a uniform
version of it.  In specific cases, for example those involving perturbations
of known endomorphism semigroups, we might expect to establish local
existence conditions of these sorts by means of fixed point theorems and
other standard methods for differential equations.

At this point, it seems worthwhile to present the motivating example for this
discussion, in the simple form of a massive scalar free field in 1+1 spacetime
dimensions.  Let $\h=\LL^2(\RR)$ be the one-particle space, and let
$\H=\exp(\h)$ be a symmetric Fock space constructed over it, whose $n$-particle
subspace $\H^n$ is the $n$-fold symmetric tensor product of $\h$ with itself.
The vectors of $\H$ we will index by the exponential map for vectors
\begin{equation}
\exp(f)=\sum^\infty_{n=0}{\vphantom\sum}^\oplus{1\over{\sqrt{n!}}}f^n
\qquad\hbox{ for $f\in\h$;}
\end{equation}
then for any $f\in\h$ we can define the unitary Weyl operator $\w(f)$ by
\begin{equation}
\w(f)\exp(g)=e^{-{1\over 2}\|f|^2-\<f| g>}\exp(f+g).
\end{equation}
If $u$ is a unitary operator on $\h$, then its multiplicative promotion $U$
given by $U\exp(f)=\exp(uf)$ will be a unitary operator on $\H$; if $a$ is a
self-adjoint operator on $\h$, then its additive promotion $A$, the generator
of the multiplicative promotion of $u(t)=e^{ita}$, will be a self-adjoint
operator on $\H$.  The additive promotion of the identity is an operator $N$,
the number operator, which has the eigenvalue $n$ on $\H^n$.  Then
for any $f\in\h$, $D_S=\cap_{n=1}^\infty D(N^n)$ will be a core
for the generator $\phi(f)$ of $\w(tf)$, such that $\phi(f)D_S\subset D_S$.

We will present $\h$ in terms of functions $f(\k)$ of the real variable $\k$,
or alternatively in terms of their Fourier transforms $\hat f(\n)$.  Then we
will let $\M$ be the von Neumann algebra generated by
$\w(f)$ for all $f$ in the real linear space
\begin{equation}
r=\[g(\k)+e^{-\pi\k}g^*(-\k)|g\in D(e^{\pi\k})].
\end{equation}
It can be shown \cite{R} that $\M'$ is the von Neumann algebra generated by
$\w(f)$ for all $f\in r'=\[g(\k)+e^{\pi\k}g^*(-\k)|g\in D(e^{-\pi\k})]$,
that $J$ is the multiplicative promotion of $j$ where $(jf)(\k)=f^*(-\k)$, and
that $\Delta^{it}$ is the multiplicative promotion of $e^{2\pi it\k}$, so that
$(\Delta^{it}\hat f)(\n)=\hat f(\n-2\pi t)$.  Then the unbounded operators
$\phi(f)$ for $f\in r$ will be self-adjoint and affiliated with $\M$,
and in fact will generate $\M$.

Clearly $h_{\lambda,\rho}=\lambda e^\n+\rho e^{-\n}$ is an unbounded
self-adjoint operator on $\h$ for each $(\lambda,\rho)\in\RR^2\setminus(0,0)$;
we may then define the self-adjoint operator $H_{\lambda,\rho}$ as the
additive promotion of $h_{\lambda,\rho}$, or alternatively by
$i[H_{\lambda,\rho},\phi(f)]=\phi(ih_{\lambda,\rho}f)$.  Then let
\begin{equation}
r_1=\[f(\k)=(\hat g(\sinh\n)+i\cosh\n\,\hat h(\sinh\n))
\check{\phantom{l}}\,|
g,h\,\,\hbox{\rm real and supported in}\,\,[1,\infty)].
\end{equation}
It can be shown that for every $(\lambda,\rho)\in\RR^2$, there is some
$\epsilon$ such that for every $f\in r_1$, $\phi(f)$ is affiliated with
$\M_\epsilon$ with respect to $H_{\lambda,\rho}$.  Furthermore $r_1+ir_1$ is
dense in $\h$.  It follows that for every $H_{\lambda,\rho}$, the local
existence condition of Theorem 2 and the uniform local existence condition
of Theorem 1 both hold.

Then for $(\lambda,\rho)\in\RR^2\setminus(0,0)$, we have the following:\\
(i) $H_{\lambda,\rho}$ is positive if and only if $\lambda$ and $\rho$ are
both non-negative;\\
(ii) $\Delta^{it}H_{\lambda,\rho}\Delta^{-it}=
H_{e^{-2\pi t}\lambda,e^{2\pi t}\rho}$, and
$JH_{\lambda,\rho}J=H_{-\lambda,-\rho}$;\\
(iii) $H_{\lambda,\rho}$ generates a one-parameter semigroup of endomorphisms
of $\M$ if and only if $\lambda$ and $-\rho$ are both non-negative; and\\
(iv) $H_{\lambda,\rho}$ generates a one-parameter semigroup of endomorphisms of
$\M'$ if and only if $-\lambda$ and $\rho$ are both non-negative.

Of course, this is a very simple example, in which it is easy to compute
the effects of the $U(a)$.  In more complicated cases, Theorems 1 and 2 could
perhaps be applied to greater effect.  However, their conditions may well be
more restrictive than is necessary; one might conjecture that the local
existence conditions could be replaced by conditions purely on the domain
$D(\delta)$ of $\delta$---for example, as in \cite{BH}, by the condition that
$D(\delta)$ be a core for $H$.

\section{Lightlike Translations}

Let us return to the situation described in the introduction, and consider
again the conditions (a)--(d).  We know already that (a) and (b) each follow
from the remaining three conditions; we have now to consider (c) and (d).
One branch is available immediately:  suppose that (a)
is satisfied, but $U(a)\Omega$ is not known.  Then
\begin{equation}
\<\Omega|U(a)\Omega\vphantom{\Delta^{it}}>
=\<\Omega|\Delta^{it}U(a)\Delta^{-it}\Omega>=
\<\Omega|U(e^{-2\pi t}a)\Omega\vphantom{\Delta^{it}}>
\end{equation}
is independent of $t$, and hence must be a constant for all $a>0$ and for all
$a<0$.  Taking the limit as $t\rightarrow\infty$, these constants must both be
$1$; but since $U(a)\Omega$ is a unit vector, it must therefore equal $\Omega$
for all $a$.  Thus (a) alone implies (d).  With this out of the way, we proceed
to our main result:

\bigskip
\noindent{\bf Theorem 3:}~~
{\em
If $H$, the generator of $U(a)$, is positive and annihilates the vacuum, and if
the local existence condition of Theorem 2 holds, then $U(a)\M U(-a)\subset\M$
for all $a\geq 0$ (and thus $U(a)\M'U(-a)\subset\M'$ for all $a\leq 0$) if and
only if the Lorentz covariance relations hold in the form
\begin{equation}
\Delta^{it}U(a)\Delta^{-it}=U(e^{-2\pi t}a)\quad{\rm and}\quad JU(a)J=U(-a).
\end{equation}
}

\smallskip
\noindent{\bf Proof of Theorem 3:}~~
{
Theorem 2 allows us to reduce this to a question about the relations
between $U(a)$ and $R$:  it will suffice to show that $U(a)R\subset R$ for
all $a\geq 0$ if and only if (8) holds.  The result of Borchers \cite{Bo}
is essentially just that (8) holds whenever $H$ is positive and
$U(a)R\subset R$ for all $a\geq 0$.  Conversely, let us assume that (8) holds.
Since $H$ is positive, $U(a)$ can be analytically continued to the upper
half-plane, and in particular we have
\begin{equation}
\Delta^t U(a)=U(a\cos(2\pi t)+ia\sin(2\pi t))\Delta^t
\end{equation}
over the region $a\geq 0$ and $0\leq t\leq 1/2$, upon which
$a\sin(2\pi t)\geq 0$.  It follows that
$U(a)D(\Delta^{1/2})\subset D(\Delta^{1/2})$ for all $a\geq 0$, and
$\Delta^{1/2}U(a)=U(-a)\Delta^{1/2}$; furthermore $JU(a)=U(-a)J$, so
that $J\Delta^{1/2}U(a)=U(a)J\Delta^{1/2}$.  But from the Tomita-Takesaki
modular theory, $R=\[\psi|\psi\in D(\Delta^{1/2}),J\Delta^{1/2}\psi=\psi]$,
so we have that $U(a)R\subset R$ for all $a\geq 0$.
}
\bigskip

Corresponding results for the backwards lightlike translations
$W(a)=T(a(\hat x_1-\hat x_0))$ can be derived by exchanging $\M$ and $\M'$,
and replacing $a$ by $-a$ in the above.  $W(a)$ should have a negative
generator, and should satisfy Lorentz covariance in the form
$\Delta^{it}W(a)\Delta^{-it}=W(e^{2\pi t}a)$ and $JW(a)J=W(-a)$.  With
these substitutions, the corresponding theorem obtains.  The situation
for the intermediate case, the spacelike translations $T(a\hat x)$ taking $W_R$
into itself, is somewhat more complicated, although not essentially different:
Theorem 1 still holds, but now the relations between the generator (which
in some frame of reference is the momentum component $P_1$) and
the modular operators are no longer so simple.  We must just show that
$T(a\hat x)D(\Delta^{1/2})\subset D(\Delta^{1/2})$ for all $a\geq 0$, and that
$J\Delta^{1/2}T(a\hat x)=T(a\hat x)J\Delta^{1/2}$.  For example, if $U(a)$
satisfies the conditions of Theorem 3, and $W(a)$ the corresponding
requirements for a backwards lightlike translation, and if $U(a)$ and $W(b)$
commute for all $a,b\in\RR$, then $U(\lambda a)W(\rho a)$ gives an endomorphism
semigroup of this intermediate type for any $\lambda,\rho>0.$  This is just the
situation described in the example at the end of the previous section.

If we combine the results of this note with those of \cite{W}, we have the
following omnibus theorem, as advertised:

\bigskip
\noindent{\bf Theorem 4:}~~
{\em Let $\M$ be a von Neumann algebra acting on a Hilbert space $\H$, which
together with its commutant $\M'$ has a separating and cyclic vector $\Omega$.
Given a strongly continuous one-parameter group $U(a)$ of unitary operators on
$\H$, for which the local existence condition of Theorem 2 holds, then
any three of the following four conditions imply the fourth:\\
(a) $\Delta^{it}U(a)\Delta^{-it}=U(e^{-2\pi t}a)$ and $JU(a)J=U(-a)$;\\
(b) the generator $H$ of the $U(a)$ is positive;\\
(c) $U(a)\M U(-a)\subset\M$ for all $a\geq 0$;\\
(d) $U(a)\Omega=\Omega$ for all $a$.\\
Likewise, any three of the following four conditions imply the fourth:\\
(a${}'$) $\Delta^{it}U(a)\Delta^{-it}=U(e^{2\pi t}a)$ and $JU(a)J=U(-a)$;\\
(b${}'$) the generator $H$ of the $U(a)$ is negative;\\
(c) $U(a)\M U(-a)\subset\M$ for all $a\geq 0$;\\
(d) $U(a)\Omega=\Omega$ for all $a$.\\
In addition, either (a) or (a${}'$) implies (d), so that if (a) holds, then
(b) and (c) are equivalent, and if (a${}'$) holds, then (b${}'$) and (c) are
equivalent; otherwise, no two of these conditions imply any other.
}


\begin{thebibliography}{99}
\bibitem{BW} J.~J.~Bisognano and E.~H.~Wichmann, J. Math. Phys. 16, 985 (1975);
                                                 J. Math. Phys. 17, 303 (1976).
\bibitem{Bo} H.~J.~Borchers, Commun. Math. Phys. 143, 315 (1992).
\bibitem{BH} O.~Bratteli and U.~Haagerup, Commun. Math. Phys. 59, 79 (1978).
\bibitem{BR} O.~Bratteli and D.~W.~Robinson, Ann. Inst. H. Poincar\'e 25A,
                                                                    139 (1976).
\bibitem{OAQSM} O.~Bratteli and D.~W.~Robinson,
{\it Operator Algebras and Quantum Statistical Mechanics}, New York:
     Springer-Verlag, 1979, Vol. I.
\bibitem{Dr} W.~Driessler, Commun. Math. Phys. 44, 133 (1975).
\bibitem{RS} M.~Reed and B.~Simon, {\it Methods of Modern Mathematical
     Physics}, Orlando:  Academic Press, 1980, Vol. I.
\bibitem{R} M.~Rieffel, Commun. Math. Phys. 39, 153 (1974).
\bibitem{W} H.~W.~Wiesbrock, Lett. Math. Phys. 25, 157 (1992).
\end{thebibliography}
\end{document}